\documentclass[a4paper,aps,pra,reprint,showkeys,floatfix,superscriptaddress]{revtex4-1}
\usepackage{graphicx}
\usepackage{amsfonts}
\usepackage{amssymb}
\usepackage{amsmath}
\usepackage{amstext}
\usepackage{amsthm}
\usepackage[percent]{overpic}

\newcommand{\Eu}{\mbox{e}}
\newcommand{\I}{\text{i}}
\newcommand{\Eg}{E_{\text{g}}}
\newcommand{\EG}{E_{\text{G}}}
\newcommand{\ket}[1]{|#1 \rangle}
\newcommand{\bracket}[2]{\langle #1|#2 \rangle}
\providecommand{\abs}[1]{\lvert#1\rvert}

\providecommand{\tfra}[2]{\tfrac{#1}{#2}}

\newcommand{\one}{\mathrm{I} \! \! 1}
\newcommand{\mbbr}{\mathbb{R}}
\newcommand{\mbbn}{\mathbb{N}}
\newcommand{\mbbz}{\mathbb{Z}}
\newcommand{\qedcube}{ }
\providecommand{\eq}[1]{Equation \eqref{#1}}
\providecommand{\fig}[1]{Figure \ref{#1}}
\providecommand{\Fig}[1]{Figure \ref{#1}}
\providecommand{\sect}[1]{Section \ref{#1}}
\newtheorem{theorem}{Theorem}
\newtheorem{lemma}[theorem]{Lemma}
\newtheorem{example}[theorem]{Example}

\begin{document}

\title{Geometric Entanglement of Symmetric States and the Majorana
  Representation}

\author{Martin Aulbach}
\email{m.aulbach1@physics.ox.ac.uk}
\affiliation{The School of Physics and Astronomy, University of Leeds,
  Leeds LS2 9JT, United Kingdom}
\affiliation{Department of Physics, University of Oxford, Clarendon
  Laboratory, Oxford OX1 3PU, United Kingdom}

\author{Damian Markham}
\email{markham@telecom-paristech.fr}
\affiliation{CNRS, LTCI, Telecom ParisTech, 23 Avenue d'Italie, 75013
  Paris, France}

\author{Mio Murao}
\email{murao@phys.s.u-tokyo.ac.jp}
\affiliation{Department of Physics, Graduate School of Science, The
  University of Tokyo, Tokyo 113-0033, Japan}
\affiliation{Institute for Nano Quantum Information Electronics, The
  University of Tokyo, Tokyo 113-0033, Japan}

\begin{abstract}
  Permutation-symmetric quantum states appear in a variety of physical
  situations, and they have been proposed for quantum information
  tasks. This article builds upon the results of [New~J.~Phys. {\bf
    12}, 073025 (2010)], where the maximally entangled symmetric
  states of up to twelve qubits were explored, and their amount of
  geometric entanglement determined by numeric and analytic means. For
  this the Majorana representation, a generalization of the Bloch
  sphere representation, can be employed to represent symmetric $n$
  qubit states by $n$ points on the surface of a unit
  sphere. Symmetries of this point distribution simplify the
  determination of the entanglement, and enable the study of quantum
  states in novel ways. Here it is shown that the duality relationship
  of Platonic solids has a counterpart in the Majorana representation,
  and that in general maximally entangled symmetric states neither
  correspond to anticoherent spin states nor to spherical designs.
  The usability of symmetric states as resources for measurement-based
  quantum computing is also discussed.  
\end{abstract}

\maketitle

\section{Introduction}\label{introduction}

Multipartite entanglement is a crucial resource for many tasks in
quantum information science, but its quantification is difficult due
to the existence of different types of entanglement \cite{Dur00}.  It
is therefore unsurprising that many different entanglement measures
have been proposed in order to quantify the amount of entanglement of
multipartite quantum states \cite{Horodecki09}.  Here we build upon
our results about highly and maximally entangled permutation-symmetric
quantum states in terms of the geometric measure of entanglement
\cite{Aulbach10}.  This restriction to a subset of quantum states --
studied under a particular entanglement measure -- makes it possible
to gain strong results \cite{Aulbach10,JMartin}, and to find a rare
visual representation of multipartite entanglement.

Permutation-symmetric quantum states are invariant under any
permutation of their subsystems.  Such states appear in many-body
physics, and they have found use in leader election
\cite{Dhondt06}. Furthermore, they have been actively implemented
experimentally \cite{Prevedel09,Wieczorek09}, and their symmetric
properties facilitate the analysis of entanglement
\cite{Bastin09,Hayashi07,Huebener,Markham10,Mathonet10,Toth09}.  In
order to analyze the usefulness of symmetric states for
measurement-based quantum computation (MBQC) \cite{VandenNest06}, the
geometric measure of entanglement is particularly suited, because the
classification of states as MBQC-resources has been performed in terms
of this measure \cite{Gross09,Mora10,VandenNest07}.

The central tool for our analysis of symmetric entanglement is the
Majorana representation \cite{Majorana32}, a generalization of the
Bloch sphere representation of single qubits. By means of this
representation any $n$ qubit symmetric state can be unambiguously
mapped to $n$ points on the surface of the unit sphere.  Recently the
Majorana representation has been used to search for and characterize
different classes of SLOCC entanglement
\cite{Aulbach11,Bastin09,Markham10,Mathonet10}, which is related to
the classification of phases in spinor condensates
\cite{Barnett07,Markham10}.  It has also been employed to search for
the ``least classical'' state of a spin-$j$ system \cite{Giraud}, and
the solutions of this problem are intimately related to the maximally
entangled symmetric states.  Furthermore, the Majorana representation
has been used for the study of spherical designs \cite{Crann10}, Berry
phases in high spin systems \cite{Hannay96}, quantum chaos
\cite{Hannay98,Leboeuf91}, optimal resources for reference frame
alignment \cite{Kolenderski08}, phase estimation \cite{Kolenderski09},
phases in spinor BEC \cite{Barnett07,Makela07}, classicality in terms
of the discriminability of states \cite{Markham03}, for finding
solutions to the Lipkin-Meshkov-Glick model \cite{Ribiero08} and for
finding efficient proofs of the Kochen-Specker theorem \cite{Zimba93}.

The article is organized as follows: In \sect{geometric_measure} we
briefly recapitulate the geometric measure of entanglement.  This is
followed by \sect{symm_states} where the geometric entanglement of
permutation-symmetric states and its implications for MBQC is
discussed.  In \sect{majorana_representation} the Majorana
representation is introduced for symmetric states of $n$ qubits, which
is followed by \sect{analytic} which reviews our analytical and
numerical findings that we recently published in \cite{Aulbach10}.  In
\sect{solutions} the usefulness of the Majorana representation is
demonstrated for highly entangled symmetric states whose point
distributions are described by Platonic solids. The entanglement of
such states is particularly easy to determine with the known
theoretical results, and it is found that there exists an intriguing
analogy with the dual polyhedra of the Platonic solids
\cite{Wenninger}.  Anticoherent spin states \cite{Zimba06} and the
mathematical concept of spherical designs \cite{Crann10} are briefly
mentioned, and it is shown that in general the maximally entangled
symmetric states do not represent anticoherent states or spherical
designs.  Finally, \sect{discussion} concludes this article with a
summary of our results.

\section{Geometric Measure of Entanglement}
\label{geometric_measure}

The geometric measure of entanglement is a distance-like entanglement
measure in the sense that it assesses the entanglement in terms of the
remoteness from the set of separable states \cite{Vedral98}.  It is
defined as the maximal overlap of a normalized pure state with all
normalized pure product states \cite{Barnum01,Shimony95,Wei03}.
\begin{equation}\label{geo_1}
  \Eg (\ket{\psi} ) = \min_{\ket{\lambda} \in
    \mathcal{H}_{\text{SEP}} } - \log_2 
    \abs{ \bracket{\lambda}{\psi} }^2 \enspace .
\end{equation}
A product state closest to $\ket{\psi}$ is denoted by
$\ket{\Lambda_{\psi}}$, and it should be kept in mind that a given
$\ket{\psi}$ can have more than one closest product state.  The
problem of maximizing the entanglement can be written as a
max-min-problem:
\begin{equation}\label{geo_meas}
  \begin{split}
    \Eg^{\text{max}} & = \max_{\ket{\psi} \in \mathcal{H}} \,
    \min_{\ket{\lambda} \in \mathcal{H}_{\text{SEP}} }
    - \log_2  \abs{\bracket{\lambda}{\psi}}^2 \\
    & = \max_{\ket{\psi} \in \mathcal{H}} - \log_2
    \abs{\bracket{\Lambda_{\psi}}{\psi}}^2 = - \log_2
    \abs{\bracket{\Lambda_{\Psi}}{\Psi}}^2 \enspace .
  \end{split}
\end{equation}

The geometric measure is closely related to the robustness of
entanglement $R$ \cite{Vidal99} and the relative entropy of
entanglement $E_{\text{R}}$ \cite{Vedral98}, two other distance-like
entanglement measures.  The inequalities $\Eg \leq E_{\text{R}} \leq
\log_2 (1 + R)$ hold for all states
\cite{Cavalcanti06,Hayashi07,Wei04}, and they become equalities for
stabilizer states, Dicke states and permutation-antisymmetric basis
states \cite{Hayashi06,Hayashi07,Markham07}.  Some advantages of the
geometric measure are its comparatively easy calculation, its
applications in related fields of physics
\cite{Lathauwer00,Nia07,Silva08}, and its operational interpretations,
e.g. in local state discrimination \cite{Hayashi06}, additivity of
channel capacities \cite{Werner02} and for the classification of
states as resources for measurement-based quantum computation
(MBQC)\cite{Gross09,Mora10,VandenNest07}.

A general quantum state of a finite-dimensional system can be cast as
$\ket{\psi} = \sum_i a_i \ket{i}$ with complex coefficients $a_i$ and
an orthonormal basis $\{ \ket{i} \}$. The state $\ket{\psi}$ is called
real if (for a given basis) the $a_i$ are all real, and positive if
the $a_i$ are all positive. Every positive state $\ket{\psi}$ has at
least one positive closest product state $\ket{\Lambda_{\psi}}$
\cite{Aulbach10,Zhu10}, a result which simplifies the determination of
their entanglement.

\section{Permutation-Symmetric States}\label{symm_states}

Permutation-symmetric quantum states are states that are invariant
under any permutation of their subsystems, i.e. $P \ket{\psi} =
\ket{\psi}$ for all $P \in S_N$.  For $n$ qubits the Hilbert space of
symmetric states is spanned by the Dicke states, the equally weighted
sums of all permutations of computational basis states with $n-k$
qubits being $\ket{0}$ and $k$ being $\ket{1}$ \cite{Dicke54,Toth03}.
\begin{equation}\label{dicke_def}
  \ket{ S_{n,k} } = {{n \choose k}}^{- 1/2} \sum_{\text{perm}} \;
  \underbrace{ \ket{0} \ket{0} \cdots \ket{0} }_{n-k}
  \underbrace{ \ket{1} \ket{1} \cdots \ket{1} }_{k} \enspace ,
\end{equation}
with $0 \leq k \leq n$.  A general pure symmetric state of $n$ qubits
is a linear combination of the $n+1$ symmetric basis states
$\ket{S_{n,k}}$.  We will abbreviate this notation to $\ket{S_{k}}$
whenever the number of qubits is clear.

It was recently found that all closest product states of multipartite
($\geq$ 3 parts) symmetric states are symmetric themselves, and that
bipartite symmetric states have at least one symmetric closest product
state \cite{Huebener}.  Furthermore, it can be shown that positive
symmetric states have at least one positive symmetric closest product
state \cite{Hayashi07}.  These results considerably reduce the
complexity of finding the closest product state and thus the
entanglement of a symmetric state.

The theoretical and experimental analysis of symmetric state
entanglement, e.g. as entanglement witnesses or in experimental setups
\cite{Korbicz05,Korbicz06,Prevedel09,Wieczorek09}, is valuable,
because symmetric states appear in many-body physics.  For example,
the ground state of the Lipkin-Meshkov-Glick model is
permutation-invariant, and its entanglement has been quantified in
term of the geometric measure \cite{Orus08}.

\subsection{Bounds on Maximal Entanglement}\label{bounds}

In this subsection we will briefly discuss the known upper and lower
bounds on the maximal possible amount of geometric entanglement. It
should however be kept in mind that the maximally entangled state and
its amount of entanglement depends on the chosen entanglement measure
\cite{Plenio07}.

The maximal possible entanglement of general $n$ qubit states scales
linearly with the number of qubits, namely
\begin{equation}\label{generalbounds}
 \tfra{n}{2}  \leq \Eg^{\text{max}} \leq n-1 \enspace .
\end{equation}
The left-hand side of the inequality is clear from the trivial example
of an $n$ qubit state ($n$ even) composed of $\frac{n}{2}$ bipartite
Bell states, or from 2D cluster states \cite{Markham07}. The upper
bound was derived in \cite{Jung08}.  It is also known that most $n$
qubit states are much closer to the upper bound than to the lower
bound.  More precisely, for $n > 10$ qubits the overwhelming majority
of states have entanglement $\Eg > n - 2 \log_2 (n) - 3$
\cite{Gross09}.

For symmetric states a trivial lower bound can be derived from the
Dicke states. A closest product state of $\ket{S_{n,k}}$ is known
\cite{Hayashi07} to be
\begin{equation}
  \label{dicke_cs}
  \ket{\Lambda} = \Big( \sqrt{ \tfra{n-k}{n} } \, \ket{0} +
  \sqrt{ \tfra{k}{n} } \, \ket{1} \Big)^{\otimes n} \enspace .
\end{equation}
From this the entanglement follows as
\begin{equation}
  \label{dicke_ent}
  \Eg ( \ket{S_{n,k}} ) = \log_2 \left( \frac{
      \big( \frac{n}{k} \big)^k \big( \frac{n}{n-k} \big)^{n-k}}
    {{n \choose k}} \right) \enspace .
\end{equation}
The maximally entangled Dicke state is $\ket{S_{n,n/2}}$ for even $n$
and the two equivalent states $\ket{S_{n, \lfloor n/2 \rfloor }}$ and
$\ket{S_{n, \lceil n/2 \rceil }}$ for odd $n$.  Their Stirling
approximation for large $n$ yields $\Eg^{\text{max}} \geq \log_2
\sqrt{n \pi/2}$.  An upper bound to the geometric measure of symmetric
$n$ qubit states has been derived from the decomposition of the
identity on symmetric subspace, yielding $\Eg^{\text{max}} \leq \log_2
(n+1)$, see e.g. \cite{Renner}. An alternative proof with the benefit
of being visually accessible by means of the Majorana representation
will be given in Theorem \ref{const_integral}.

Combining these bounds, it is seen that the maximal symmetric
entanglement of $n$ qubits scales as
\begin{equation}\label{symmetricbounds}
  \log_2 \sqrt{\tfra{n \pi}{2}} \leq \Eg^{\text{max}} \leq
  \log_2 (n+1) \enspace ,
\end{equation}
i.e. polylogarithmically between $\mathcal{O} (\log \sqrt{n})$ and
$\mathcal{O} (\log n)$. Numerical evidence suggests that the actual
values are much closer to the upper bound than to the lower bound, and
$\Eg^{\text{max}} \gtrsim \log_2 (n+1) - 0.775$ can be considered a
reliable lower bound \cite{JMartin}.

\subsection{Resources for MBQC}
\label{resources_for_mbqc}

We have seen that the maximal entanglement of symmetric states scales
much slower than that of general states, namely logarithmically rather
than linearly. This need not be a disadvantage for symmetric states,
though, and in fact could render them useful for MBQC
\cite{VandenNest06}, because it was shown that if the entanglement of
a state is too large, then it cannot be a good resource for MBQC. More
specifically, if the $n$ qubit entanglement scales larger than $n -
\delta$ for some constant $\delta$, then such a computation can be
simulated efficiently classically \cite{Gross09}.  This rules out many
general quantum states as MBQC resources, but not symmetric ones.

On the other hand, universal resources for MBQC must be maximally
entangled in a certain sense \cite{Mora10,VandenNest07}. Considering
the qualitative departure of the scaling relation
\eqref{symmetricbounds} from \eqref{generalbounds}, it is questionable
whether symmetric states are sufficiently entangled to be MBQC
resources.  Indeed, permutation-symmetric states can be ruled out as
exact, deterministic MBQC resources, because their entanglement does
not scale faster-than-logarithmically
\cite{Aulbach10,VandenNest07}. Somewhat weaker requirements are
imposed upon approximate, stochastic MBQC resources \cite{Mora10},
although this generally leads only to a small extension of the class
of suitable resources in the vicinity of exact, deterministic
resources (e.g. 2D cluster states with holes).  It is therefore
believed that symmetric states cannot be used even for approximate,
stochastic MBQC.

As an example, we will show that Dicke states with a fixed number of
excitations cannot be useful for $\epsilon$-approximate, deterministic
MBQC \cite{Mora10}. Roughly speaking, $\epsilon$-approximate universal
resource states can be converted into any other state by LOCC with an
inaccuracy of at most $\epsilon$.  The $\epsilon$-version of the
geometric measure \cite{NoteGeoMeasDef} is defined as \cite{Mora10}
\begin{equation}\label{epsilon_entanglement_definition}
  \EG^{\epsilon} (\rho) = \min \{ \EG (\sigma)
  \, \vert \, D(\rho,\sigma) \leq \epsilon \} \enspace ,
\end{equation}
where $D$ is a distance that is ``strictly related to the fidelity'',
meaning that for any two states $\rho$ and $\sigma$, $D(\rho,\sigma)
\leq \epsilon \Rightarrow F(\rho,\sigma) \geq 1 - \eta (\epsilon)$,
where $0 \leq \eta (\epsilon) \leq 1$ is a strictly monotonically
increasing function with $\eta (0) = 0$.  $\EG^{\epsilon} (\rho)$ can
be understood as the guaranteed entanglement obtained from a
preparation of $\rho$ with inaccuracy $\epsilon$. One possible choice
of $D$ is the trace distance, which for pure states reads
$D_{\text{t}} (\ket{\psi},\ket{\phi}) = \sqrt{1 -
  \abs{\bracket{\psi}{\phi}}^2 } = \sqrt{1 - F}$, where $F$ is the
fidelity. In this case one can choose $\eta(\epsilon) = \epsilon^{2}$.

As shown in Example 1 of \cite{Mora10}, the family of W states
$\Psi_{\text{W}} = \{ \ket{\mbox{W}_{n} } \}_{n}$, with
$\ket{\mbox{W}_{n} } \equiv \ket{ S_{n,1} }$, is not an
$\epsilon$-approximate universal resource for $\eta(\epsilon) \lesssim
0.001$.  This result can be generalized to all families of Dicke
states $\Psi_{{S}_{k}} = \{ \ket{S_{n,k} } \}_{n}$ with a fixed number
of excitations $k$.

\begin{example}
  For any fixed $k \in \mbbn$ the family of Dicke states
  $\Psi_{{S}_{k}} = \{ \ket{S_{n,k} } \}_{n}$ cannot be an
  $\epsilon$-approximate universal MBQC resource for $\eta(\epsilon)
  \lesssim 0.001 \, k^{-3/2}$.
\end{example}
\begin{proof}
  Using \eq{dicke_ent} and the Stirling approximation for high $n$,
  the asymptotic geometric entanglement of the family $\Psi_{{S}_{k}}$
  is found to be
\begin{equation}\label{asy_ent}
  \EG ( \Psi_{{S}_{k}} ) = 1 - \frac{k^k}{\Eu^{k} k!} \enspace .
\end{equation}
Specifically, the amount of geometric entanglement remains finite for
arbitrary values of $n$, allowing us to apply Proposition 3 and
Theorem 1 of \cite{Mora10} to show that the necessary condition for
$\epsilon$-approximate deterministic universality,
\begin{equation}\label{approx_cond}
  \EG ( \Psi_{{S}_{k}} ) > 1-4 \eta^{1/3} + 3.4 \eta^{2/3} \enspace ,
\end{equation}
is violated for $\eta(\epsilon) \lesssim 0.001 \, k^{-3/2}$. \qedcube
\end{proof}

Of course, it should be noted that many other quantum information
tasks are not restricted by the requirements of MBQC-universality, and
that highly entangled symmetric states can therefore be valuable
resources for such tasks.

\section{Majorana Representation of Symmetric States}
\label{majorana_representation}

The classical angular momentum $\mathbf{J}$ of a physical system can
be represented by a single point on the surface of the unit sphere in
$\mbbr^3$, corresponding to the direction of $\mathbf{J}$.  Quantum
mechanics does not allow for such a simple representation, but it is
possible to uniquely represent a pure state of spin-$j$ by $2j$
undistinguishable points on the sphere \cite{Majorana32}.  This is a
generalization of the Bloch sphere representation of a qubit.  An
equivalent representation can be shown to exist for symmetric states
of $n$ spin-$(1/2)$ particles \cite{Bacry74,Majorana32}, with an
isomorphism mediating between all states of a spin-$j$ particle and
the symmetric states of $2j$ qubits.

Hence, this ``Majorana representation'' allows us to uniquely compose
any symmetric state of $n$ qubits $\ket{\psi}_{\text{s}}$ from a sum
over all permutations $P : S_N \rightarrow S_N$ of $n$
undistinguishable qubits $\{ \ket{\phi_1} , \dots , \ket{\phi_n} \}$:
\begin{gather}
  \ket{\psi}_{\text{s}} = K^{- 1/2} \sum_{ \text{perm} }
  \ket{\phi_{P(1)}} \ket{\phi_{P(2)}} \cdots \ket{\phi_{P(n)}}
  \enspace \mbox{, with}
  \label{majorana_definition} \\
  \ket{\phi_i} = \cos \tfra{\theta_i}{2} \, \ket{0} + \Eu^{\I
    \varphi_i} \sin \tfra{\theta_i}{2} \ket{1} \enspace , \nonumber
\end{gather}
and where the normalization factor $K$ depends on the given state.
The identity \eqref{majorana_definition} allows the visualization of
the multi-qubit state $\ket{\psi}_{\text{s}}$ by $n$ points on a
sphere. In the following these points will be called the
\emph{Majorana points} (MP), and the sphere on which they lie the
\emph{Majorana sphere}.

As outlined in the previous section, for $n \geq 3$ qubits every
closest product state $\ket{\Lambda}$ of a symmetric state
$\ket{\psi}_{\text{s}}$ is symmetric itself \cite{Huebener}, and can
therefore be written as $\ket{\Lambda} = \ket{\sigma}^{\otimes n}$,
with a single qubit state $\ket{\sigma}$.  The closest product states
of a given symmetric state can therefore be visualized by Bloch
vectors too, and in analogy to the Majorana points, we call
$\ket{\sigma}$ a \emph{closest product point} (CPP).

For symmetric states the scalar product from the definition of the
geometric measure can be concisely expressed in terms of the MPs and a
CPP:
\begin{equation}\label{bloch_product}
  \abs{ \bracket{\Lambda}{\psi}_{\text{s}} } =
  n! \, K^{- 1/2} \, \prod_{i=1}^{n} \,
  \abs{ \bracket{\sigma}{\phi_i}} \enspace .
\end{equation}
To determine the CPP of a given symmetric state, one therefore has to
maximize the absolute value of a product of scalar products. The
factors $\bracket{\sigma}{\phi_i}$ are the angles between the
corresponding Bloch vectors on the Majorana sphere, thus turning the
determination of the CPP into a geometrical optimization problem.

From \eq{majorana_definition} it follows that the application on an
arbitrary single-qubit unitary operation $U$ to each of the $n$
subsystems of a symmetric state $\ket{\psi}_{\text{s}}$ yields
\begin{equation}\label{lusphere}
  \begin{split}
    \ket{\varphi}_{\text{s}} & = U^{\otimes n} \ket{\psi}_{\text{s}} \\
    & = K^{- 1/2} \sum_{\text{perm}} \! \left( U \ket{\phi_{P(1)}}
    \right) \otimes \cdots \otimes \left( U \ket{\phi_{P(n)}} \right)
    \enspace .
  \end{split}
\end{equation}
Thus the symmetric state $\ket{\psi}_{\text{s}}$ is mapped to a
symmetric state $\ket{\varphi}_{\text{s}}$ whose MP distribution can
be obtained from a joint rotation of the MPs of
$\ket{\psi}_{\text{s}}$ along a common axis on the Majorana sphere.
The two LOCC-equivalent states $\ket{\psi}_{\text{s}}$ and
$\ket{\varphi}_{\text{s}}$ have the same \emph{relative} MP
distribution, and therefore the same number and relative distribution
of CPPs, as well as the same amount of entanglement.

\subsection{Examples}\label{examples}

\begin{figure}
  \centering
  \begin{overpic}[scale=.7]{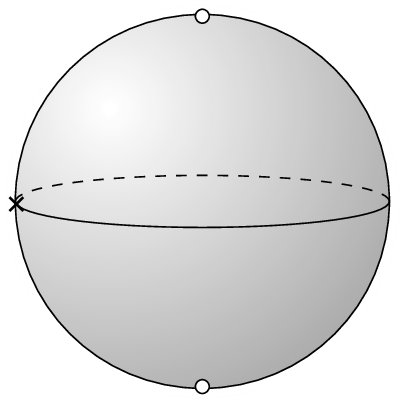}
    \put(-2,0){(a)}
    \put(35,83){$\ket{\phi_1}$}
    \put(36,10){$\ket{\phi_2}$}
    \put(8,58){$\ket{\sigma_1}$}
  \end{overpic}
  \begin{overpic}[scale=0.7]{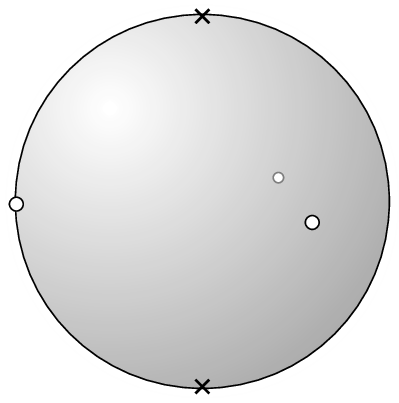}
    \put(-2,0){(b)}
    \put(8,45){$\ket{\phi_1}$}
    \put(72,32){$\ket{\phi_2}$}
    \put(62,62){$\ket{\phi_3}$}
    \put(35,83){$\ket{\sigma_1}$}
    \put(36,10){$\ket{\sigma_2}$}
  \end{overpic}
  \begin{overpic}[scale=0.7]{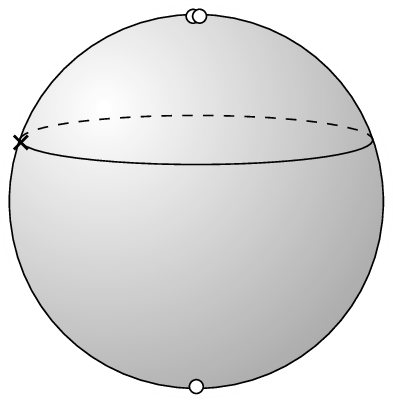}
    \put(-2,0){(c)}
    \put(32,83){$\ket{\phi_1}$}
    \put(52,83){$\ket{\phi_2}$}
    \put(36,10){$\ket{\phi_3}$}
    \put(11,67){$\ket{\sigma_1}$}
  \end{overpic}
  \caption{\label{bell_ghz_w} Majorana representations of symmetric
    states of two and three qubits.  MPs are depicted as white dots
    and CPPs as crosses or dashed lines. The pictures show (a) the two
    qubit Bell state $\ket{\psi^{+}}$, (b) three qubit GHZ state and
    (c) three qubit W state.}
\end{figure}

For pure symmetric states of two qubits the only absolute degree of
freedom in the Majorana representation (and hence entanglement) is the
distance between the two MPs.  It is easy to determine that the CPP
lies halfway between the two MPs, and that the entanglement is
maximized when the MPs lie antipodal to each other.  \Fig{bell_ghz_w}
(a) shows the Bell state $\ket{\psi^{+}} = 1/\sqrt{2} \left( \ket{01}
  + \ket{10} \right)$ with its two MPs $| \phi_1 \rangle = |0\rangle$
and $| \phi_2 \rangle = |1\rangle$.  Due to this azimuthal symmetry
the CPPs form a continuous ring $\ket{\sigma} = 1 / \sqrt{2} \left(
  \ket{0} + \Eu^{\I \varphi} \ket{1} \right)$, with $\varphi \in [0,2
\pi)$ around the equator.  The amount of entanglement is $\Eg (
\ket{\psi^{+}} ) = 1$.

For three qubits the GHZ state and W state, two positive symmetric
states, are considered to be extremal \cite{Tamaryan}, with the W
state proven to be the maximally entangled state in terms of the
geometric measure \cite{Chen}.

The MPs of the tripartite GHZ state $\ket{\mbox{GHZ}} = 1 / \sqrt{2}
\left( \ket{000} + \ket{111} \right)$ are, up to normalization,
\begin{equation}\label{GHZ-maj}
  \begin{split}
    \ket{\phi_1} & = \ket{0} + \ket{1} \enspace , \\
    \ket{\phi_2} & = \ket{0} + \Eu^{\I 2 \pi / 3} \ket{1} \enspace , \\
    \ket{\phi_3} & = \ket{0} + \Eu^{\I 4 \pi / 3} \ket{1} \enspace .
  \end{split}
\end{equation}
Its two CPPs are $\ket{\sigma_1} = \ket{0}$ and $\ket{\sigma_2} =
\ket{1}$, and the amount of entanglement is $\Eg ( \ket{\mbox{GHZ}} )
= 1$. \Fig{bell_ghz_w}(b) shows the Majorana representation of the GHZ
state.  The three MPs form an equilateral triangle on the equator, and
the two CPPs are the north pole and south pole.

In the case of the W state $\ket{\mbox{W}} = \ket{S_{3,1}} = 1 /
\sqrt{3} \left( \ket{001} + \ket{010} + \ket{100} \right)$, a Dicke
state, the MPs can be directly accessed from its definition as
$\ket{\phi_1} = \ket{\phi_2} = \ket{0}$ and $\ket{\phi_3} = \ket{1}$.
The positive CPP follows from \eq{dicke_cs} as $\ket{\sigma_1} =
\sqrt{ 2/3 } \, \ket{0} + \sqrt{ 1/3 } \, \ket{1}$, and the azimuthal
symmetry implies that the set of all CPPs is formed by $\ket{\sigma} =
\sqrt{2/3} \, \ket{0} + \Eu^{\I \varphi} \sqrt{1/3} \, \ket{1}$, with
$\varphi \in [0,2 \pi)$. The Majorana representation is shown in
\fig{bell_ghz_w}(c), and the entanglement is $\Eg ( \ket{\mbox{W}} ) =
\log_2 \left( 9/4 \right) \approx 1.17$.

\subsection{Extremal Point Distributions}
\label{extremal_point}

With \eq{bloch_product} the min-max-problem \eqref{geo_meas} of
finding the maximally entangled symmetric state can be recast as
\begin{equation}\label{maj_problem}
  \min_{ \{ \ket{\phi_i} \}} K^{- 1/2} \left( \max_{ \ket{\sigma} } \,
    \prod_{i=1}^{n} \, \abs{ \bracket{\sigma}{\phi_i} } \right)
  \enspace .
\end{equation}
Solving this ``Majorana problem'' is far from trivial, particularly
with the normalization factor $K$ depending on the MPs.  The problem
can be understood as an optimization problem on the sphere, prompting
the question whether the known solutions of classical point
distribution problems on the sphere \cite{Whyte52} can help in finding
the solutions of the Majorana problem.  Two problems that have been
extensively studied in the past are T\'{o}th's problem and Thomson's
problem.

T\'{o}th's problem states that $n$ points have to be distributed over
the sphere so that the minimum pairwise distance becomes maximal
\cite{Whyte52}. Point configurations that solve this problem are known
as spherical codes or sphere packings.

Thomson's problem is considering $n$ point charges which are confined
to the surface of a sphere and interacting with each other through
Coulomb's inverse square law. The desired distribution is the one
which minimizes the potential energy \cite{Thomson04}.  This problem
has a variety of applications, e.g. for multi-electron bubbles in
liquid Helium \cite{Leiderer95}, liquid metal drops confined in Paul
traps \cite{Davis97}, shell structure of viruses \cite{Marzec93},
colloidosomes \cite{Dinsmore02}, fullerene patterns \cite{Kroto85} and
Abrikosov lattice of vortices in superconducting metal shells
\cite{Dodgson97}.

\begin{figure}[b]
  \centering
  \begin{overpic}[scale=0.83]{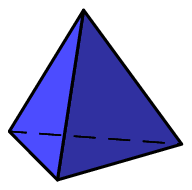}
  \end{overpic}
  \begin{overpic}[scale=0.83]{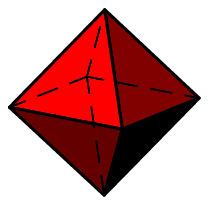}
  \end{overpic}
  \hspace{0.1mm}
  \begin{overpic}[scale=0.78]{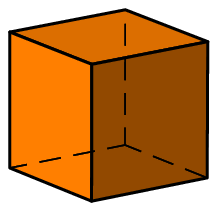}
  \end{overpic}
  \hspace{0.1mm}
  \begin{overpic}[scale=0.77]{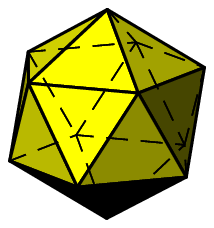}
  \end{overpic}
  \begin{overpic}[scale=0.80]{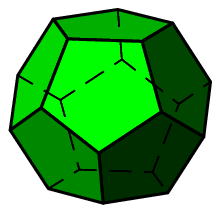}
  \end{overpic}
  \caption{\label{platonic} (color online) The five Platonic solids
    from left to right: tetrahedron ($n=4$), octahedron ($n=6$), cube
    ($n=8$), icosahedron ($n=12$), and dodecahedron ($n=20$).}
\end{figure}

Exact solutions to T\'{o}th's and Thomson's problem of $n$ points are
known only for very few and low $n$ \cite{Erber91,Whyte52}, but
numerical solutions are known for a much wider range of $n$ in both
problems \cite{Altschuler94,Ashby86}.  An illustrating example are the
five Platonic solids -- the regular convex polyhedra whose edges,
vertices and angles are all congruent, see \fig{platonic}.  Because of
their high symmetry one would expect that their vertices solve
T\'{o}th's and Thomson's problem for the corresponding $n$. This is
however true only for $n=4,6,12$, but not for $n = 8,20$.

\begin{figure}
  \centering
  \begin{overpic}[scale=1.0]{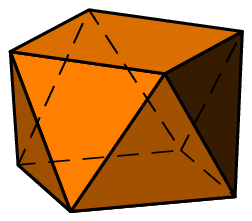}
  \end{overpic}
  \hspace{5mm}
  \begin{overpic}[scale=0.8]{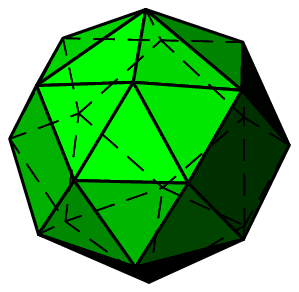}
  \end{overpic}
  \caption{\label{toth} (color online) For $n=8$ the solution of
    T\'{o}th's problem is given by a cubic antiprism, and for $n=20$
    by a polyhedron consisting of 30 triangles and 3 rhombuses.}
\end{figure}

\Fig{toth} depicts the polyhedra that solve T\'{o}th's problem for $n
= 8,20$. For $n=8$ the solution is the cubic antiprism, which can be
obtained from the cube by rotating one face by 45 degrees, followed by
a slight compression along the direction perpendicular to the rotated
face. In this way, the nearest neighbor distances between the vertices
can be equally reduced, at the expense of breaking the high Platonic
symmetry.  This simple example shows that it is in general not easy to
find the optimal spherical distribution for a set of points, and this
is also true for the Majorana problem.

\section{Analytic Results about MPs and CPPs}
\label{analytic}

This section summarizes the analytic results about the Majorana
representation that we have presented in \cite{Aulbach10}.  In
particular, the relationship between the coefficients of a symmetric
state $\ket{\psi}_{\text{s}} = \sum_{k = 0}^{n} a_k \ket{S_{k}}$ and
the corresponding distribution of MPs and CPPs on the Majorana sphere
will be illuminated.

\begin{theorem}\label{const_integral}
  For every symmetric $n$ qubit state $\ket{\psi}_{\text{s}}$ the
  following holds:
  \begin{equation}\label{intvolume}
    \int\limits_{0}^{2 \pi} \int\limits_{0}^{\pi}
    \abs{\bracket{\lambda(\theta , \varphi)}{\psi}_{\text{s}}}^2
    \sin \theta \, \emph{d} \theta \emph{d} \varphi =
    \frac{4 \pi}{n+1} \enspace ,
  \end{equation}
  where $\ket{\lambda(\theta , \varphi)} = \left( \cos
    \tfra{\theta}{2} \ket{0} + \emph{e}^{\emph{i} \varphi} \sin
    \tfra{\theta}{2} \ket{1} \right)^{\otimes n}$.
\end{theorem}
For the proof of this theorem we refer to \cite{Aulbach10}.  The
remarkable property of \eq{intvolume} is that the integral is the same
for all symmetric $n$ qubit states, thus straightforwardly yielding
the upper bound $\Eg^{\text{max}} \leq \log_2 (n+1)$ on the maximal
symmetric entanglement.  The integrand of \eq{intvolume} can be
visualized by a spherical plot, and the constant integration volume
can be understood as the constant volume of the plot.
\Fig{icosahedronpic1}(b) shows such a plot for a symmetric 12 qubit
state.

Majorana representations with a high degree of symmetry are
particularly easy to investigate.  It is therefore elucidating to know
the necessary and sufficient conditions for a rotational symmetry of
the MP distribution.

\begin{lemma}\label{rot_symm}
  The MP distribution of a symmetric $n$ qubit state
  $\ket{\psi}_{\text{s}}$ is rotationally symmetric around the
  $Z$-axis with rotational angle $\theta = \tfra{2 \pi}{m}$ ($\, 1<m
  \leq n$) iff
\begin{equation}\label{rot_cond}
  \forall \{ k_i , k_j | \, a_{k_{i}} \neq 0 \wedge
  a_{k_{j}} \neq 0 \} : ( k_i - k_j ) \bmod m = 0 \enspace .
\end{equation}
\end{lemma}
This lemma states that all non-vanishing coefficients must be spaced
apart from each other by a multiple of $m>1$. An example of a
rotationally symmetric state with $\theta = \pi / 2$ would be
$\ket{\psi}_{\text{s}} = a_3 \ket{S_3} + a_7 \ket{S_7} + a_{15}
\ket{S_{15}}$.

Symmetric states whose coefficients are all real can be associated
with a reflective symmetry of the Majorana representation along the
$X$-$Z$-plane.  From a mathematical point of view two Bloch vectors
$\ket{\phi_1}$ and $\ket{\phi_2}$ exhibit such a reflective symmetry
iff they are complex conjugates, i.e. $\ket{\phi_1} = \cos
\tfra{\theta}{2} \ket{0} + \Eu^{\I \varphi} \sin \tfra{\theta}{2}
\ket{1}$ and $\ket{\phi_2} = \cos \tfra{\theta}{2} \ket{0} + \Eu^{- \I
  \varphi} \sin \tfra{\theta}{2} \ket{1} = \ket{\phi_1}^{*}$.

\begin{lemma}\label{maj_real}
  Let $\ket{\psi}_{\text{s}}$ be a symmetric state of $n$ qubits.
  $\ket{\psi}_{\text{s}}$ is real iff all its MPs are reflective
  symmetric with respect to the $X$-$Z$-plane of the Majorana sphere.
\end{lemma}
It immediately follows from the form of the min-max-problem
\eqref{maj_problem} that this reflective symmetry is also inherited to
the CPPs.

Particularly strong results about the number and locations of CPPs can
be obtained for positive symmetric states.  With the exception of the
Dicke states, any positive symmetric state can have at most $2n-4$
CPPs, and it is believed that this result also holds for general
symmetric states.  Dicke states are a special case due to their
continuous azimuthal symmetry, resulting in an uncountable number of
CPPs.

\begin{lemma}\label{cpp_mer}
  Let $\ket{\psi}_{\text{s}}$ be a positive symmetric state of $n$
  qubits, excluding the Dicke states.

  \begin{enumerate}
  \item[(a)] If $\ket{\psi}_{\text{s}}$ is not rotationally symmetric
    around the Z-axis, then all its CPPs are positive.

  \item[(b)] If $\ket{\psi}_{\text{s}}$ is rotationally symmetric
    around the Z-axis with minimal rotational angle $\tfra{2 \pi}{m}$,
    then all its CPPs $\ket{\sigma (\theta, \varphi)} = \cos
    \tfra{\theta}{2} \ket{0} + \emph{e}^{\emph{i} \varphi} \sin
    \tfra{\theta}{2} \ket{1}$ are restricted to the $m$ azimuthal
    angles given by $\varphi = \varphi_{r} = \tfra{2 \pi r}{m}$ with
    $r \in \mbbz$.  Furthermore, if $\ket{\sigma (\theta, \varphi_{r}
      ) }$ is a CPP for some $r$, then it is also a CPP for all other
    values of $r$.
  \end{enumerate}
\end{lemma}
The restriction of the CPPs to certain azimuthal angles imposed by
this lemma is crucial for the rather technical proof (c.f. Appendix B
of \cite{Aulbach10}) of the following statement about the number and
locations of the CPPs.

\begin{theorem}\label{maj_max_pos_zero}
  The Majorana representation of every positive symmetric state of $n$
  qubits, excluding the Dicke states, belongs to one of the following
  three classes.

  \begin{enumerate}
  \item[(a)] $\ket{\psi}_{\text{s}}$ is rotationally symmetric around
    the Z-axis, with only the two poles as possible CPPs.

  \item[(b)] $\ket{\psi}_{\text{s}}$ is rotationally symmetric around
    the Z-axis, with at least one CPP being non-positive.

  \item[(c)] $\ket{\psi}_{\text{s}}$ is not rotationally symmetric
    around the Z-axis, and all CPPs are positive.
  \end{enumerate}
  Regarding the CPPs of states from class (b) and (c), the following
  assertions can be made for $n \geq 3$ qubits:
  \begin{enumerate}
  \item[(b)] If both poles are occupied by at least one MP each, then
    there are at most $2n-4$ CPPs, else there are at most $n$ CPPs.

  \item[(c)] There are at most $\lceil \tfra{n+2}{2} \rceil$ CPPs
  \end{enumerate}
\end{theorem}
The upper bound on the number of CPPs is intriguing, because the Euler
characteristic implies that convex polyhedra with $n$ vertices have at
most $2n-4$ faces.  One could therefore ask whether there exists a
deeper relationship between the CPPs and the faces of the MP
distribution.

\section{Solutions for up to Twelve Qubits}
\label{solutions}

An exhaustive search for the maximally entangled symmetric state over
the whole space of symmetric states becomes infeasible already for
only a few qubits, because the min-max-problem \eqref{maj_problem} is
too intractable to easily determine solutions.  The results from the
previous section as well as the fact that the maximally entangled
state must have at least two CPPs (c.f. Lemma 4 in \cite{Aulbach10})
considerably simplify the numerical search for high and maximal
symmetric entanglement, particularly among the subset of positive
symmetric states, allowing the reliable determination of the maximally
entangled positive symmetric states of up to 12 qubits.  For the
general non-positive case an exhaustive search over the entire Hilbert
space is still too involved, so we concentrated on sets of promising
states. Such states include those with highly spread out MP
distributions and those that share qualitative features with the
solutions to the classical optimization problems.  Table
\ref{enttable} summarizes the presumed values of maximal geometric
entanglement for symmetric states in the positive and general
case. For comparison purposes, the known upper and lower bounds are
also listed.  For a detailed presentation and discussion of all the
solutions we refer to \cite{Aulbach10}.

\begingroup
\squeezetable
\begin{table}
  \centering
  \caption{\label{enttable} Values for the maximal entanglement of
    symmetric $n$ qubit states in terms of the geometric measure.
    The entanglement values listed are (from left to right) those of
    the most entangled Dicke state, the maximally entangled positive
    symmetric state, the presumably maximally entangled symmetric
    state and the upper bound on symmetric entanglement.  The
    relation $\Eg \!  \left( \ket{S_{{\lfloor n/2 \rfloor}}} \right)
    \leq \Eg \!  \left( \ket{\Psi^{\text{pos}}_{n}} \right) \leq \Eg
    \!  \left( \ket{\Psi_{n}} \right) < \log_2 (n + 1)$ holds for
    all $n$, and wherever the amount of entanglement does not
    increase, the respective right-hand table cell has been
    intentionally left blank.  All numerical values have been
    calculated for ten or more digits, and the dagger $\dagger$ in
    the second column indicates values whose analytic form is known,
    but not displayed due to their complicated form.}
  \begin{tabular}{lcccc}
    \hline\noalign{\smallskip}
    $n$ & $\Eg \! \left( \ket{S_{{\lfloor n/2 \rfloor}}} \right)$ &
    $\Eg \! \left( \ket{\Psi^{\text{pos}}_{n}} \right)$
    & $\Eg \! \left( \ket{\Psi_{n}} \right)$ & $\log_2 (n+1)$ \\
    \noalign{\smallskip}
    \hline
    \noalign{\smallskip}
    2 & $1$ & & & $\log_2 3$ \\
    3 & $\log_2 (9/4)$ & & & $2$ \\
    4 & $\log_2 (8/3)$ & $\log_2 3$ & & $\log_2 5$ \\
    5 & $1.532 \: 824 \: 877$ &
    $1.742 \: 268 \: 948 \!$ $^\dagger$ & &
    $2.584 \: 962 \: 501$ \\
    6 & $\log_2 (16/5)$ & $\log_2 (9/2)$ & & $\log_2 7$ \\
    7 & $1.767 \: 313 \: 935$ &
    $2.298 \: 691 \: 396 \!$ $^\dagger$ & & $3$ \\
    8 & $1.870 \: 716 \: 983$ &
    $2.445 \: 210 \: 159 \!$ \phantom{$^\dagger$} & &
    $3.169 \: 925 \: 001$ \\
    9 & $1.942 \: 404 \: 615$ &
    $2.553 \: 960 \: 277 \!$ $^\dagger$  & &
    $3.321 \: 928 \: 095$ \\
    10 & $2.022 \: 720 \: 077$ &
    $2.679 \: 763 \: 092 \!$ \phantom{$^\dagger$} &
    $2.737 \: 432 \: 003$ & $3.459 \: 431 \: 619$ \\
    11 & \phantom{i} $2.082 \: 583 \: 285$ \phantom{i} &
    $2.773 \: 622 \: 669 \!$ \phantom{$^\dagger$} &
    \phantom{i} $2.817 \: 698 \: 505$ \phantom{i} &
    $3.584 \: 962 \: 501$ \\
    12 & $2.148 \: 250 \: 959$ &
    $2.993 \: 524 \: 700 \!$ \phantom{$^\dagger$} &
    $\log_2 (243/28)$ & $3.700 \: 439 \: 718$ \\
    \hline
  \end{tabular}
\end{table}
\endgroup

For $n=2,3$ qubits the maximally entangled states were already
identified as the Bell states and the W state, respectively.  For
$n=4,6,12$ the Majorana problem is solved by the respective Platonic
solid, i.e. the MP distributions are given by the vertices of the
corresponding Platonic solid.

\begin{figure}[b]
  \centering
  \begin{minipage}{80mm}
    \begin{overpic}[scale=.43]{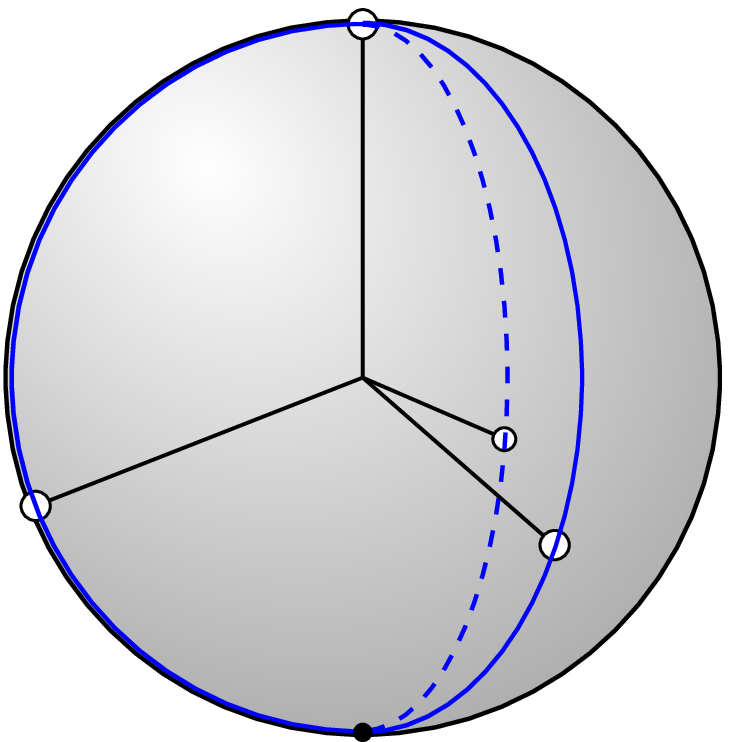}
      \put(-2,0){(a)}
      \put(41,103){$| \phi_1 \rangle$}
      \put(-12,20){$| \phi_2 \rangle$}
      \put(77,31){$|  \phi_3 \rangle$}
      \put(51,53){$|  \phi_4 \rangle$}
    \end{overpic}
    \hspace{5mm}
    \begin{overpic}[scale=.43]{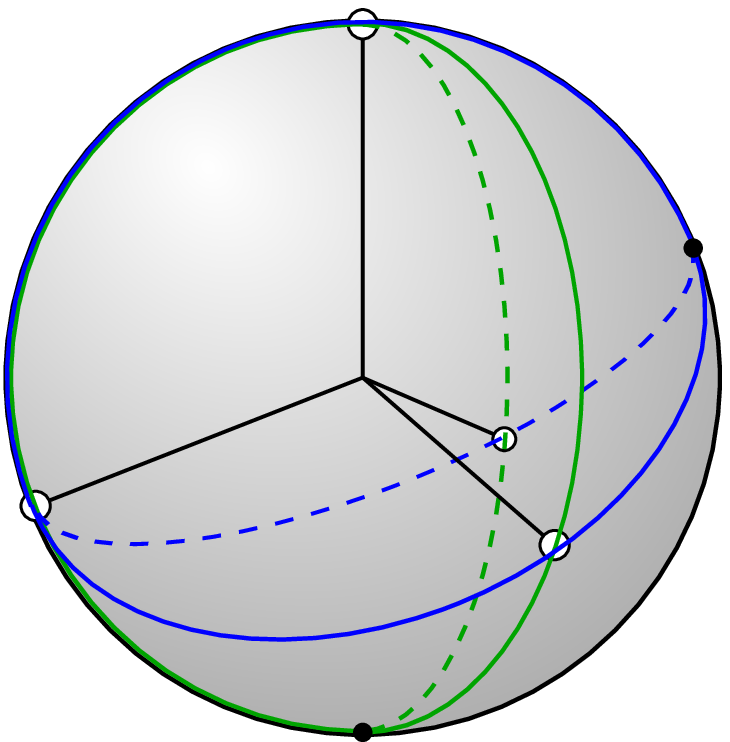}
      \put(-2,0){(b)}
      \put(41,103){$|  \phi_3 \rangle$}
      \put(-12,20){$| \phi_1 \rangle$}
      \put(77,31){$|  \phi_2 \rangle$}
      \put(51,53){$|  \phi_4 \rangle$}
    \end{overpic}
    \vspace{5mm}
  \end{minipage}
  \begin{minipage}{80mm}
    \begin{overpic}[scale=.43]{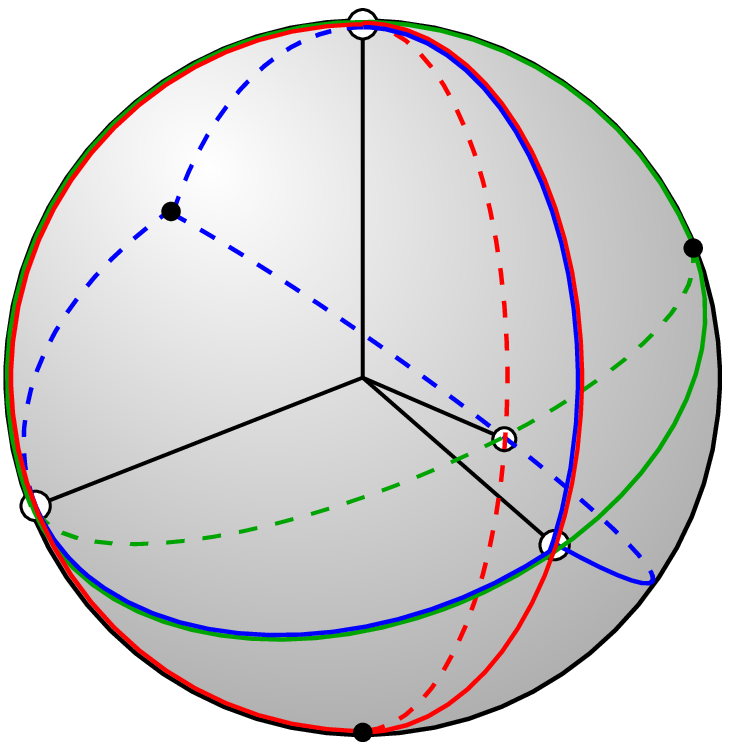}
      \put(-2,0){(c)}
      \put(41,103){$|  \phi_2 \rangle$}
      \put(-12,20){$| \phi_3 \rangle$}
      \put(77,31){$|  \phi_1 \rangle$}
      \put(51,53){$|  \phi_4 \rangle$}
    \end{overpic}
    \hspace{5mm}
    \begin{overpic}[scale=.82]{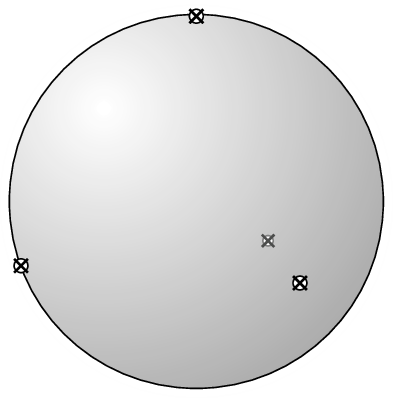}
      \put(-2,0){(d)}
    \end{overpic}
  \end{minipage}
  \caption{\label{4_cpps} (color online) The CPPs of the positive
    symmetric tetrahedron state $\ket{\Psi_4}$ of four qubits can be
    directly obtained from the tetrahedral rotation group and Lemma
    \ref{cpp_mer}.  Applying finite rotations amounts to permutations
    of the MPs and thus additional restrictions for the locations of
    the CPPs are obtained from Lemma \ref{cpp_mer}.}
\end{figure}

The ``tetrahedron state'' of four qubits, shown in \fig{4_cpps}, has
the form $\ket{\Psi_{4}} = 1 / \sqrt{3} \, \ket{S_{0}} + \sqrt{2/3} \,
\ket{S_{3}}$. Since the state is positive and has a Z-axis rotational
symmetry, Lemma \ref{cpp_mer} restricts the CPPs to the three
half-circles shown as blue lines in \fig{4_cpps}(a).  By means of the
tetrahedral rotation group it is possible to find a unitary operation
$U \neq \one$ so that \eq{lusphere} maps $\ket{\Psi_{4}}$ onto
itself. This can be understood as a rotation on the Majorana sphere
which moves each MP to the location of another MP.  A rotation of this
type, with the Bloch vector of $\ket{\phi_4}$ acting as the rotation
axis, is performed twice between \fig{4_cpps}(a) and \fig{4_cpps}(c).
For each of these configurations Lemma \ref{cpp_mer} gives rise to
separate restrictions on the locations of the CPPs, and the
intersection of all these restrictions leaves only four points, the
MPs themselves. Therefore $| \Psi_{4} \rangle$ has four CPPs which
coincide with the MPs.

\begin{figure}
  \centering
  \begin{overpic}[scale=.37]{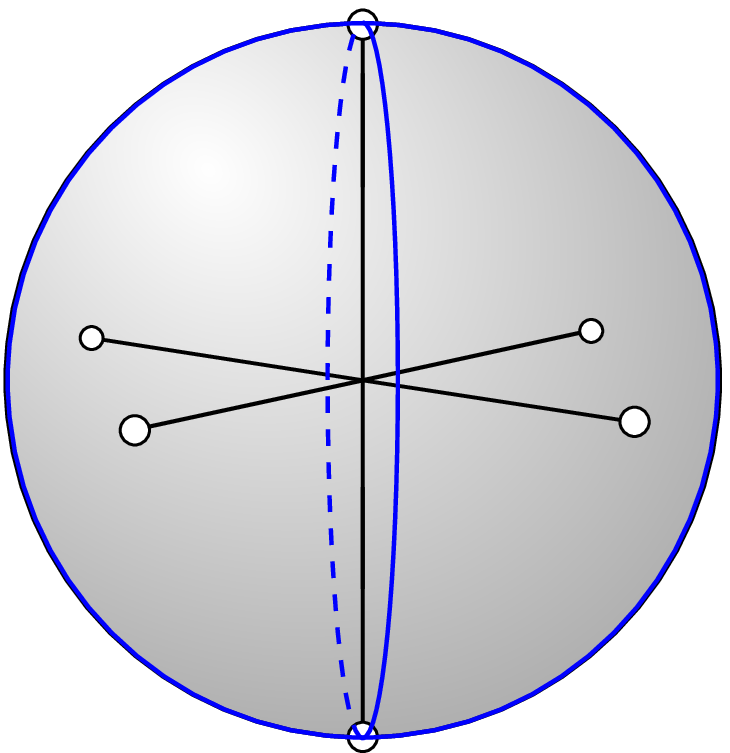}
    \put(-2,0){(a)}
    \put(35,104){$\ket{\phi_1}$}
    \put(35,-9){$\ket{\phi_2}$}
    \put(9,30){$\ket{\phi_3}$}
    \put(65,32){$\ket{\phi_4}$}
    \put(68,66){$\ket{\phi_5}$}
    \put(10,63){$\ket{\phi_6}$}
  \end{overpic}
  \hspace{0.11mm}
  \begin{overpic}[scale=.37]{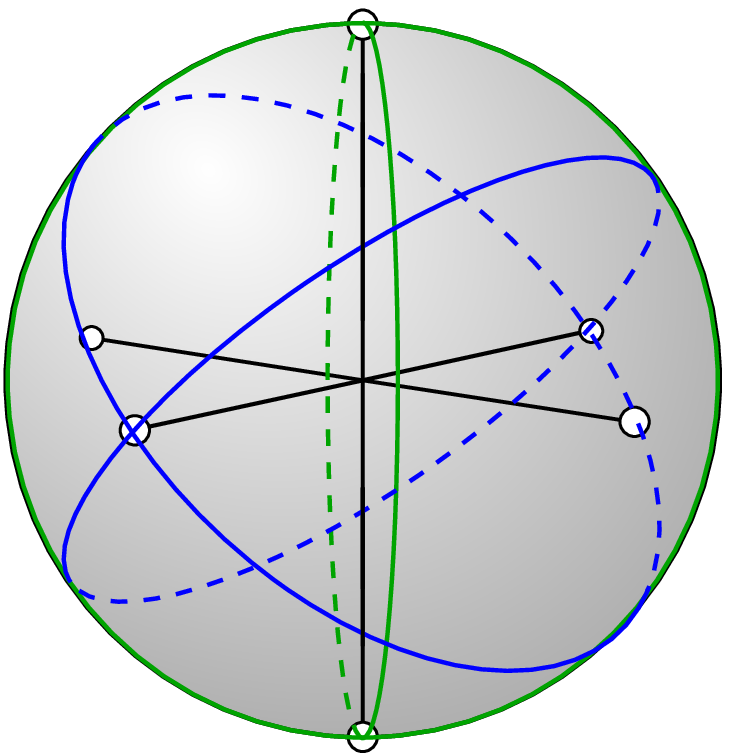}
    \put(-2,0){(b)}
    \put(35,104){$\ket{\phi_5}$}
    \put(35,-9){$\ket{\phi_3}$}
    \put(9,30){$\ket{\phi_1}$}
    \put(65,32){$\ket{\phi_4}$}
    \put(68,66){$\ket{\phi_2}$}
    \put(10,63){$\ket{\phi_6}$}
  \end{overpic}
  \hspace{0.01mm}
  \begin{overpic}[scale=.712]{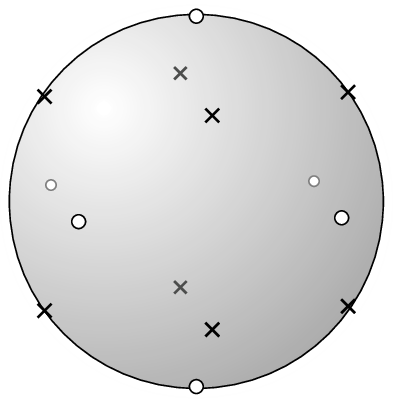}
    \put(0,0){(c)}
  \end{overpic}
  \caption{\label{6_cpps} (color online) Only one rotation from the
    octahedral rotation group is required to uniquely determine the
    locations of the eight CPPs of the octahedron state
    $\ket{\Psi_6}$.}
\end{figure}

For the ``octahedron state'' of six qubits $\ket{\Psi_{6}} = 1 /
\sqrt{2} ( \ket{S_{1}} + \ket{S_{5}} )$, shown in \fig{6_cpps}, the
CPPs can be determined in the same way. Only one rotation from the
octahedral rotation group is required to find the eight CPPs at the
intersections of the blue and green lines depicted in
\fig{6_cpps}(b). The CPPs lie at the center of each face of the
octahedron, forming a cube inside the Majorana sphere.  In contrast to
the tetrahedron state with its overlapping MPs and CPPs, the CPPs of
the octahedron state lie as far away from the MPs as possible.  This
is because the expression \eqref{maj_problem} would be zero if a CPP
$\ket{\sigma}$ were to lie antipodal to a MP $\ket{\phi_i}$.

\begin{figure}[b]
  \centering
  \begin{overpic}[scale=.82]{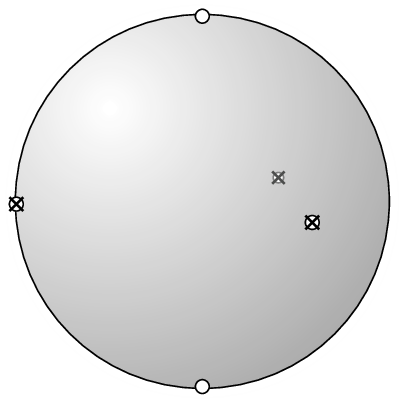}
    \put(-2,0){(a)}
    \put(41,103){$\ket{\phi_1}$}
    \put(41,-7){$\ket{\phi_2}$}
    \put(-15,54){$\ket{\phi_3}$}
    \put(74,32){$\ket{\phi_4}$}
    \put(60,62){$\ket{\phi_5}$}
  \end{overpic}
  \hspace{5mm}
  \begin{overpic}[scale=.82]{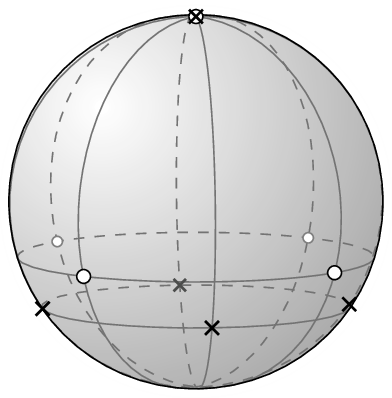}
    \put(-2,0){(b)}
    \put(41,103){$\ket{\phi_1}$}
    \put(23,21){$\ket{\phi_2}$}
    \put(64,22){ $\ket{\phi_3}$}
    \put(62,46){$\ket{\phi_4}$}
    \put(4,45){$\ket{\phi_5}$}
  \end{overpic}
  \caption{\label{bloch_5} The ``trigonal bipyramid state'' is shown
    in (a), but the Majorana problem of five qubits is solved by the
    ``square pyramid state'' shown in (b).}
\end{figure}

For five points the solution to the classical problems is the trigonal
bipyramid \cite{Ashby86}, and the corresponding ``trigonal bipyramid
state'' $\ket{\psi_{5}} = 1 / \sqrt{2} ( \ket{S_{1}} + \ket{S_{4}} )$
is shown in \fig{bloch_5}(a).  This is however not the maximally
entangled symmetric state, and a numerical search yields the ``square
pyramid state'' $\ket{\Psi_{5}} \approx 0.547 \ket{S_{0}} + 0.837
\ket{S_{4}}$, shown in \fig{bloch_5}(b), as the maximally entangled
one.  All its MPs and CPPs can be determined analytically by solving
quartic equations.  One on the five CPPs coincides with the north pole
while the other four are equidistantly spread over a horizontal plane
in the southern hemisphere.  Notably, the ``center of mass'' of the
five MPs of $\ket{\Psi_5}$ does not coincide with the origin of the
sphere, and the implications of this will be outlined in
\sect{anticoherent_queens}.

\begin{figure}
  \centering
  \begin{overpic}[scale=0.9]{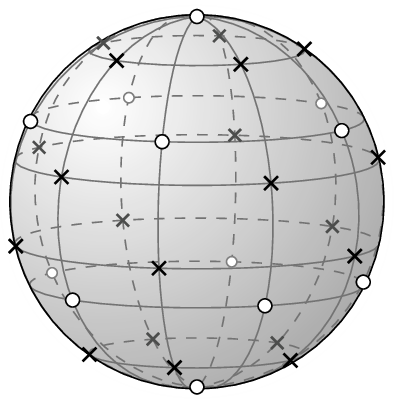}
    \put(-5,0){(a)}
  \end{overpic}
  \hspace{5mm}
  \begin{overpic}[scale=.12]{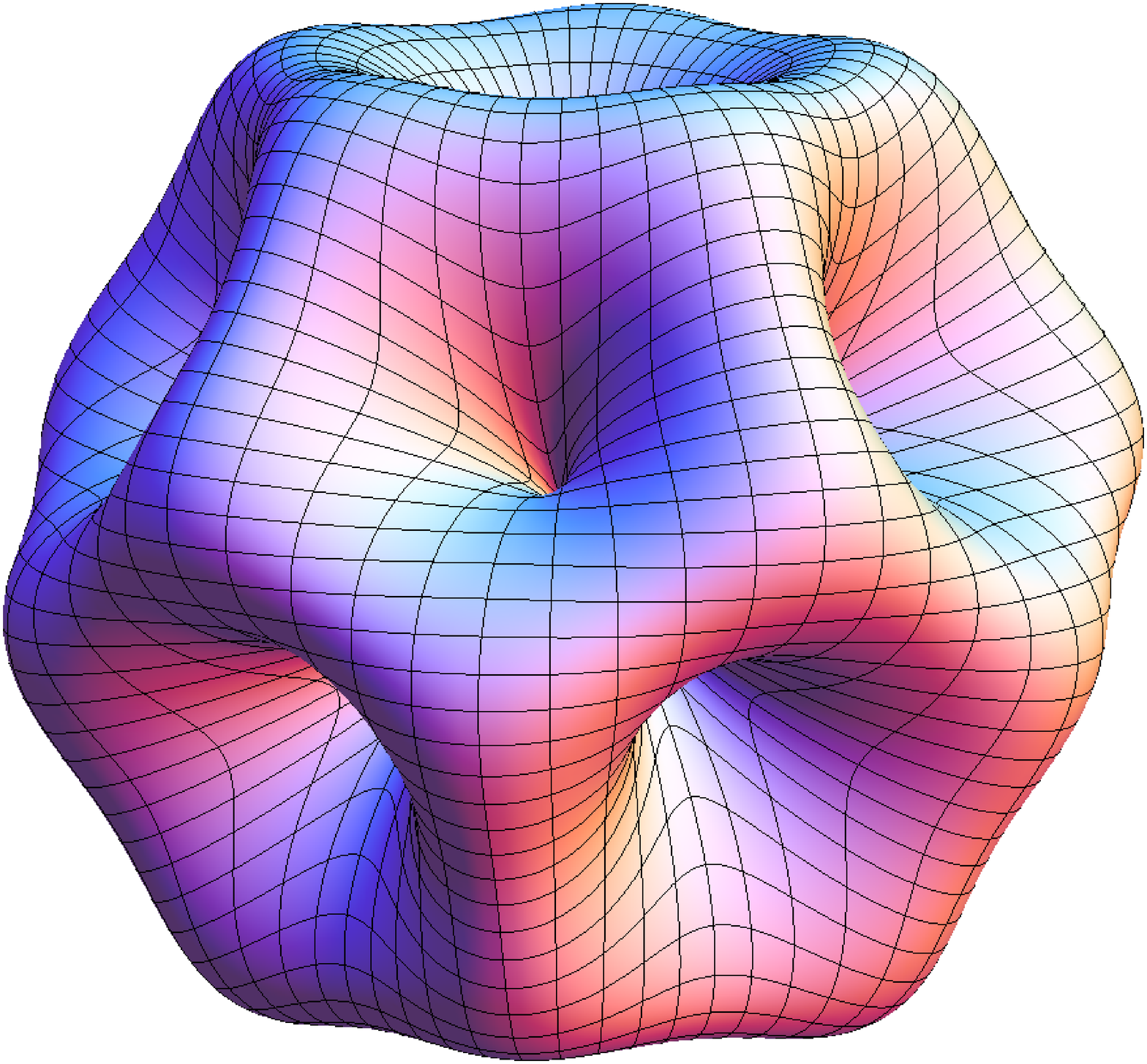}
    \put(-5,0){(b)}
  \end{overpic}
  \caption{\label{icosahedronpic1} (color online) The MPs and CPPs of
    the 12 qubit ``icosahedron state'' $\ket{\Psi_{12}}$ are depicted
    in (a), and the corresponding amplitude function $f(\theta,
    \varphi) = \abs{\bracket{\lambda(\theta , \varphi)}{\Psi_{12}}}$
    is shown in (b). For $\ket{\Psi_{12}}$ the locations of the MPs
    and CPPs coincide with the zeros and maxima of $f(\theta,
    \varphi)$, respectively.}
\end{figure}

There is strong evidence that the ``icosahedron state''
$\ket{\Psi_{12}} = \sqrt{7} \, \ket{S_{1}} - \sqrt{11} \, \ket{S_{6}}
- \sqrt{7} \, \ket{S_{11}}$, shown in \fig{icosahedronpic1}(a), is the
maximally entangled symmetric state of 12 qubits. The MPs form the
vertices of a regular icosahedron, while the 20 CPPs are centered on
the faces of the icosahedron, describing a dodecahedron inside the
Majorana sphere.  \Fig{icosahedronpic1}(b) is the spherical plot of
the function $f(\theta, \varphi) = \abs{\bracket{\lambda(\theta ,
    \varphi)}{\Psi_{12}}}$ which already appeared as the integrand of
\eq{intvolume}.  This function is variously known as the
characteristic polynomial, Majorana polynomial \cite{Kolenderski08},
amplitude function \cite{Radcliffe71} or coherent state decomposition
\cite{Leboeuf91}. The CPPs and MPs of a symmetric state can be readily
identified as the global maxima and the antipodes of the zeros of
$f(\theta, \varphi)$, respectively.

\subsection{Dual Polyhedra}

\begin{figure}[b]
  \centering
  \begin{overpic}[scale=1.2]{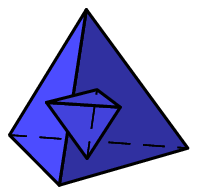}
  \end{overpic}
  \hspace{2mm}
  \begin{overpic}[scale=1.1]{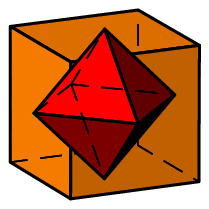}
  \end{overpic}
  \hspace{2mm}
  \begin{overpic}[scale=1.1]{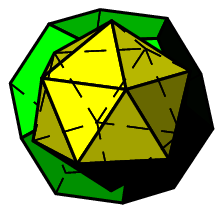}
  \end{overpic}
  \caption{\label{platonic_dual} (color online) The relationships
    between the Platonic solids and their duals.}
\end{figure}

Each of the five Platonic solids shown in \fig{platonic} has a dual
polyhedron with faces and vertices interchanged, and this dual
polyhedron is again a Platonic solid \cite{Wenninger}. As seen in
\fig{platonic_dual}, the octahedron and cube form a dual pair, and so
do the icosahedron and dodecahedron. In contrast to this, the
tetrahedron is self-dual, i.e. it is its own dual.

Interestingly, these dualities are also inherited to the Majorana
representations of the corresponding symmetric quantum states. For
example, we have seen that the 20 CPPs of the icosahedron state
$\ket{\Psi_{12}}$ form the vertices of a dodecahedron. On the other
hand, when considering the 20 qubit ``dodecahedron state''
$\ket{\Psi_{20}} = \sqrt{187} \ket{S_{0}} + \sqrt{627} \ket{S_{5}} +
\sqrt{247} \ket{S_{10}} - \sqrt{627} \ket{S_{15}} + \sqrt{187}
\ket{S_{20}}$, it is easy to show that this state has 12 CPPs which
occupy the vertices of an icosahedron. Thus the Majorana
representation of the dodecahedron state can be immediately obtained
from \fig{icosahedronpic1} (a) by interchanging the MPs and CPPs.  The
same duality exists between the octahedron state and the cube state,
c.f.  \fig{6_cpps} (c). Furthermore, the tetrahedron state is its own
dual, as seen in \fig{4_cpps} (d). Unlike the dual of the Platonic
solid, however, the dual tetrahedron state is not turned ``upside
down'' as seen in \fig{platonic_dual}, but rather coincides with the
original tetrahedron state.

\subsection{Anticoherent Spin States and the Queens of
  Quantum}\label{anticoherent_queens}

As outlined in \sect{majorana_representation}, there exists an
isomorphism between the states of a spin-$j$ particle and the
symmetric states of $2j$ qubits.  The coherent states of a quantum
particle can be regarded as the most classical states, and in terms of
the Majorana representation these states are those whose MPs all
coincide at a single point, thus describing a ``classical'' spin
vector.  Anticoherent spin states, first studied in \cite{Zimba06},
are states that exhibit maximally nonclassical behavior in the sense
that their spin vector vanishes.  Since such states can be considered
the ``opposite'' of coherent states, it would be interesting to
determine the MPs and the geometric entanglement of their symmetric
counterparts. For example, one could ask whether maximally entangled
symmetric states correspond to anticoherent states or to the
mathematical concept of spherical designs \cite{Crann10}.  However,
the fact that the ``center of mass'' of the five qubit square pyramid
state $\ket{\Psi_{5}}$ does not coincide with the origin of the
Majorana sphere straightforwardly implies that this state is neither
anticoherent nor a spherical design \cite{Notecenterofmass}.

An alternative to anticoherent states was formulated in \cite{Giraud},
where the least classical states are coined ``queens of quantum''.
The Majorana representations of these states differ from our maximally
entangled symmetric states, but when replacing the Hilbert-Schmidt
metric with the Bures metric \cite{JMartin}, the solutions of the two
problems become identical.  In other words, the Majorana
representation of the spin-$j$ ``queen of quantum'' in terms of the
Bures metric is identical to that of the maximally entangled symmetric
state of $2j$ qubits in terms of the geometric measure.

\section{Conclusion}\label{discussion}

We have analyzed and discussed the geometric entanglement of highly
and maximally entangled symmetric states of $n$ qubits.  The upper
bound on symmetric entanglement rules out symmetric states as exact,
deterministic MBQC resources.  For the case of approximate MBQC we
present arguments against the usefulness of symmetric states, and
provide a proof for the class of Dicke states.  With the known
analytic results about the Majorana representation of symmetric states
it is easy to numerically determine the most entangled states and to
discuss their properties. As an example we showed how the
determination of the CPPs of ``Platonic states'' is greatly simplified
with the help of the theoretical results. With the help of the
maximally entangled symmetric five qubit state it was shown that the
solutions to the Majorana problem do not necessarily relate to
anticoherent states or spherical designs.  It is found that the
well-known concept of the dual polyhedra of Platonic solids possesses
a direct analog for symmetric quantum states, thereby deepening the
relationship between the Majorana representation and the polyhedra of
classical geometry.

\begin{acknowledgments}
  The authors would like to thank S.~Miyashita, A.~Soeda, S.~Virmani,
  K.-H.~Borgwardt and M.~Van~den~Nest for very helpful discussions.
  This work is supported by the National Research Foundation \&
  Ministry of Education, Singapore and the project ``Quantum
  Computation: Theory and Feasibility'' in the framework of the
  CNRS-JST Strategic French-Japanese Cooperative Program on ICT. MM
  thanks the ``Special Coordination Funds for Promoting Science and
  Technology'' for financial support.
\end{acknowledgments}


\begin{thebibliography}{10}
\bibitem{NoteGeoMeasDef}
There exist two common definitions of the geometric measure, which we
  distinguish by denoting $\Eg (\ket{\psi}) = - \log_2
  \abs{\bracket{\Lambda_{\psi}}{\psi}}^2$ and $\EG (\ket{\psi}) = 1 -
  \abs{\bracket{\Lambda_{\psi}}{\psi}}^2$. The former is used throughout this
  article, with the only exception being this subsection, where $\EG$ is used
  in order to be consistent with the notation of \cite{Mora10}.
\bibitem{Notecenterofmass}
For anticoherent spin states this readily follows from $\bracket{\Psi_{5}}{S_z
  \vert \Psi_{5}} \neq 0$. For spherical designs we observe that by setting
  $p(x) = x$ in Definition 2 of \cite{Crann10}, it follows that for all
  spherical designs the ``center of mass'' must necessarily coincide with the
  sphere's origin.
\end{thebibliography}
\end{document}